\documentclass[floatfix,%
reprint,
superscriptaddress,
 amsmath,amssymb,
 aps,
prc,
]{revtex4-2}

\usepackage{graphicx}
\usepackage{dcolumn}
\usepackage{bm}
\usepackage{soul} 
\usepackage[mathlines]{lineno}
\usepackage{color}

\usepackage{braket}

\newcommand{\nucl}[2]{{}^{#1}\mathrm{#2}}

\newcommand{\ob}{\mathrm{ob}}
\newcommand{\pr}{\mathrm{pr}}

\newcommand{\BE}[2]{B(\mathrm{E{#1}}; #2)}
\newcommand{\coeffdep}{c_3}

\begin{document}

\preprint{}

\title{Oblate-prolate shape mixing and E0 transition in $\bm{\nucl{28}{Si}}$}

\author{Yasutaka Taniguchi}
 \affiliation{Department of Computer Science, Fukuyama University}
 \email{taniguchi-y@fukuyama-u.ac.jp}
 \affiliation{RIKEN Nishina Center}
\author{Masaaki Kimura}%
\affiliation{RIKEN Nishina Center}%

\date{\today}

\begin{abstract}
\begin{description}
\item[Background] 
oblate-prolate shape coexistence in $^{28}$Si has been discussed for decades, but the degree of shape mixing between these configurations remains poorly constrained.
\item[Purpose]
We constrain the oblate-prolate mixing amplitudes in $^{28}$Si using available experimental information and  discuss the inter-band E0 transition strength.
\item[Methods] 
Oblate and prolate $0^+$ and $2^+$ configurations are obtained by antisymmetrized molecular dynamics combined with the generator coordinate method. Using these configurations as the basis states, we constrain the mixing amplitudes by simultaneously reproducing the measured charge radius, the quadrupole moment of the $2_1^+$ state, and the in-band and inter-band $B(E2)$ values. The strength of the density-dependent term in the Gogny interaction is also varied within a reasonable range.
\item[Results]
In the ground state, the oblate component is dominant, and the prolate component in the ground state is limited to less than about $20\%$. For the $2_1^+$ state, the allowed prolate component is smaller than that in the ground state. The present analysis does not tightly constrain the corresponding E0 transition strength, but an upper limit of
$\rho^2(E0;0_3^+\rightarrow0_1^+) \lesssim 0.206$ is obtained.
\item[Conclusions]
The low-lying $0^+$ states of $^{28}$Si may exhibit substantial oblate-prolate mixing. A measurement of the inter-band $E0$ transition strength would provide a quantitative determination of the mixing amplitude.
\end{description}
\end{abstract}
\maketitle

\section{Introduction}
Static quadrupole deformation is one of the most successful concepts in nuclear physics. It provides a simple interpretation of rotational spectra and enhanced electromagnetic transitions observed in many nuclei~\cite{Bohr1976,BenderHeenenReinhard2003,ProchniakRohozinski2009,NiksicVretenarRing2011}. However, nuclei are finite quantum systems, and their shapes are not necessarily rigid. Collective wave functions generally show fluctuations in deformation, and when different shapes have close energies, substantial mixing between them can occur~\cite{HeydeWood2011,GarrettZielinskaClement2022}. Quantifying such shape mixing is an interesting problem for understanding the low-lying collective states.

The $\nucl{28}{Si}$ nucleus is a famous example of oblate-prolate shape coexistence~\cite{Bernier1967,gupta1967,DoDang1970,PhysRevC.71.014303,taniguchi,chiba,Frycz2024}.
The positive quadrupole moment of the $2_1^+$ state $(0.16\pm0.03~\mathrm{b})$~\cite{Stone2016}
indicates that the ground band is predominantly oblate.
Although there is no direct experimental evidence,
many HF~\cite{gupta1967,DoDang1970} and AMD~\cite{PhysRevC.71.014303,taniguchi,chiba} calculations consistently suggest a prolate deformation for the $K^\pi=0_3^+$ band built on the $0_3^+$ state.
An important question is whether these two configurations coexist independently or are mixed by quantum fluctuations.
The most direct probe of shape mixing is the inter-band E0 transition strength~\cite{HeydeWood2011},
but it has not yet been measured.
However, the observed finite inter-band E2 transition strength, $\BE{2}{0_3^+\rightarrow2_1^+}=1.35\pm0.10~\mathrm{fm}^4$~\cite{Basunia2013}, implies shape mixing.
Other observables, such as the in-band E2 transition strengths, the charge radius, and the quadrupole moment, are also affected by shape mixing.
Combined with microscopic structure calculations, they potentially provide constraints on the oblate-prolate mixing amplitudes and may allow an estimation of the inter-band E0 transition strength.

Dang~\cite{DoDang1970} performed a pioneering study of oblate-prolate shape mixing and demonstrated that the observed inter-band $0_3^+ \to 2_1^+$ and in-band $2_1^+ \to 0_1^+$ transition strengths can be explained by introducing an ad-hoc interaction that induces the mixing of the oblate and prolate shapes. 
However, since the model space was limited to the $sd$-shell, it could not describe different radii for the oblate and prolate configurations, and hence, the inter-band E0 transition strength was not discussed. 
Subsequent studies employing more realistic interactions and larger model spaces, including shell-model calculations in the $sdpf$ shell~\cite{Frycz2024} and our previous AMD calculations~\cite{taniguchi,chiba}, described the coexistence of the oblate and prolate shapes. 
However, these calculations failed to reproduce the observed $\BE{2}{0_3^+\rightarrow2_1^+}$ value. 
Thus, the oblate-prolate mixing amplitudes remain poorly constrained.

In this work, we take a different approach to the problem. Rather than predicting the mixing amplitudes from a given effective interaction, we treat them as free parameters and constrain them using the available experimental data. To this end, we use AMD wave functions of the oblate and prolate configurations as microscopic basis states and construct their mixed states with varying amplitudes. The allowed mixing amplitudes are then determined by simultaneously reproducing the observed inter-band and in-band E2 transition strengths, the charge radius of the ground state, and the quadrupole moment of the $2_1^+$ state within their experimental uncertainties. The resulting constraints also allow us to estimate the inter-band E0 transition strength $\rho^2(\mathrm{E0};0_3^+\rightarrow0_1^+)$.

This paper is organized as follows.
In Sec.~\ref{sec:framework}, we describe the AMD+GCM calculation used to generate the oblate and prolate basis states and show that a direct configuration-mixing calculation underestimates the observed inter-band E2 transition strength.
In Sec.~\ref{sec:constraint}, we introduce the procedure for constraining the mixing amplitudes from the charge radius, quadrupole moment, and E2 transition strengths and determine the allowed parameter region.
We also estimate the inter-band E0 transition strength within the allowed parameter region.
Finally, we summarize this work.

\section{AMD+GCM calculation for basis generation}\label{sec:framework}
In this section, we generate oblate- and prolate-dominant basis states using the AMD+GCM framework. We first describe the framework and then discuss the properties of the resulting low-lying states. Although the AMD+GCM calculation reproduces the coexistence of the oblate and prolate-deformed rotational bands, the inter-band transition strengths remain much smaller than the experimental values due to insufficient oblate-prolate mixing. These oblate- and prolate-dominant states are used as basis states for the mixing-amplitude analysis presented in the following section.

\subsection{Framework of AMD+GCM}
An AMD wave function $\Phi^\pi$ is a parity-projected Slater determinant of the single-particle wave packets, whose spatial part is described by a deformed Gaussian wave packet~\cite{Kimura2004}.
\begin{align}
    \Phi^\pi &= P^\pi\mathcal{A}\{\varphi_1, \varphi_2, ..., \varphi_A\},
    \quad \pi=\pm\\
    \varphi_i &= e^{
      -(\mathbf{r} - \mathbf{Z}_i) \cdot \mathsf{M}(\mathbf{r}-\mathbf{Z}_i)}
      \otimes (a_i \chi_{\uparrow} + b_i\chi_{\downarrow}) \otimes \eta_i,
\end{align}
where $P^\pi$ is the parity projection operator, $\chi_{\uparrow}$ and $\chi_{\downarrow}$ are the spin-up and down wave functions, while $\eta_i$ is the isospin wave function fixed to either a proton or a neutron.
The Gaussian centroid $\mathbf{Z}_i$ is the complex vector, and the width matrix $\mathsf{M}$ is a  real positive-definite symmetric matrix common to all nucleons.
The spin direction is parametrized by the complex parameters $a_i$ and $b_i$.

First, we perform constrained variational calculations with $(\beta,\gamma)$ constraints to generate AMD basis wave functions with
various axial prolate and oblate deformations.
Then, the $J^\pi=0^+$ and $2^+$ states, which are relevant to the shape-mixing problem discussed below, are obtained by the GCM calculations,
\begin{equation}
    \Ket{J_n^{+}} = \sum_{iK} f_{niK} P_{MK}^{J}\ket{\Phi^\pi(\beta_i)},
\end{equation}
where $P^{J}_{MK}$ is the angular momentum projection operator, and the coefficient $f_{niK}$ is determined by the Hill-Wheeler equation. 

As the effective nucleon-nucleon interaction, we use the Gogny D1S interaction~\cite{DechargeGogny1980,Berger1991} with a scaling parameter $c_3$ introduced for the density-dependent term,
\begin{align}
    \coeffdep t_3 (1+P_\sigma)\left[\rho\left(\frac{\mathbf{r}_1 + \mathbf{r}_2}{2}\right)\right]^{1/3}\delta(\mathbf{r}_1-\mathbf{r}_2).\label{eq:gogny}
\end{align}
The motivation for this modification will be discussed in Sec.~\ref{sec:underestimate}.

\subsection{Results of the AMD+GCM calculation}\label{sec:underestimate}
\begin{figure}
    \centering
    \includegraphics[width=\linewidth]{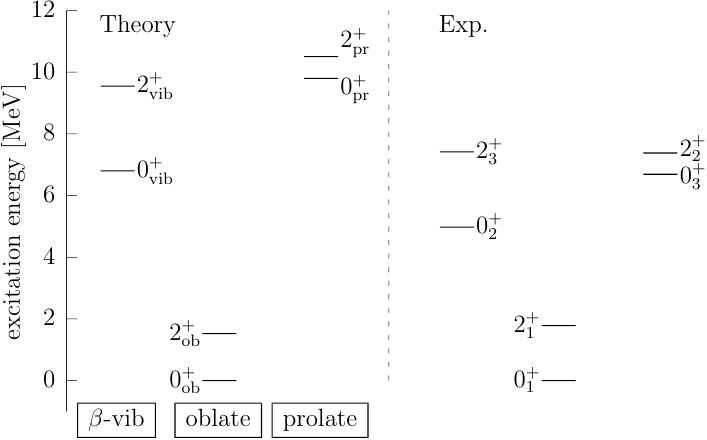}
    \caption{
    Theoretical and experimental level scheme of low-lying $0^+$ and $2^+$ states. 
    }
    \label{fig:level}
\end{figure}

\begin{table}[tbp]
    \centering
    \caption{Comparison of observables of the AMD+GCM and experimental data. Experimental data are taken from Refs.~\onlinecite{Angeli2013,Stone2016,Basunia2013}.}
    \label{tab:gcm}
\begin{ruledtabular}
\begin{tabular}{ccc}
        observable &  AMD+GCM & Exp.\\
        \hline
        charge radius ($0_1^+$) [fm] & 3.19 & $3.1224 \pm 0.0024$ \\
        Q-moment ($2_1^+$) [b]& $+0.17$ & $+0.16 \pm 0.03$\\
        $\BE{2}{2_1^+ \to 0_1^+}~[\mathrm{fm}^4]$ & 72.7 & $66.6 \pm 2.5$\\
        $\BE{2}{0_3^+ \to 2_1^+}~[\mathrm{fm}^4]$ & $<10^{-3}$ & $1.35\pm 0.10$\\
        $\rho^2(\mathrm{E0}; {0_3^+ \to 0_1^+})$ & $<10^{-5}$ & --\\
    \end{tabular}
    \end{ruledtabular}
\end{table}

Figure~\ref{fig:level} compares the calculated and experimental low-lying $0^+$ and $2^+$ states, 
and Table~\ref{tab:gcm} summarizes the charge radii, quadrupole moments, and E0 and E2 transition strengths.
The calculated level scheme consists of three groups. The $0_\ob^+$ and $2_\ob^+$ states, which are dominated by the oblate deformed configuration, correspond to the observed $0^+_1$ and $2^+_1$ states. The positive quadrupole moment of the $2_{\ob}^+$ state and the large in-band $\BE{2}{2_{\ob}^+ \rightarrow 0_{\ob}^+}$ reproduce the observed value, supporting their oblate deformation. On the other hand, the $0_\pr^+$ and $2_\pr^+$ states are dominated by a prolate configuration, and we consider that they correspond to the observed $0^+_3$ and $2^+_3$ states. Although direct experimental evidence is unavailable, this assignment is consistent with previous theoretical studies~\cite{DoDang1970,PhysRevC.71.014303,taniguchi,chiba,Frycz2024}.
The $0_{\rm vib}^+$ and $2_{\rm vib}^+$ states are interpreted as $\beta$-vibrational states~\cite{taniguchi,chiba} rather than oblate–prolate mixed states and are assigned to the observed $0^+_2$ and $2^+_2$ states. Since the present work focuses on the mixing between the oblate- and prolate-dominant bands, these states are excluded from the following analysis.

Despite the reasonable description of the oblate and prolate deformed states, the calculated inter-band transition strength $\BE{2}{0_\pr^+\rightarrow2_\ob^+}$ is much smaller than the corresponding observed value. The E0 matrix element between the $0^+_{\ob}$ and $0^+_{\pr}$ states is also extremely small. Since the inter-band transitions are sensitive to shape mixing, these results indicate that the AMD+GCM wave functions contain only a negligible amount of oblate–prolate mixing. Therefore, the states $0_{\rm ob}^+$, $2_{\rm ob}^+$, $0_{\rm pr}^+$, and $2_{\rm pr}^+$ may be regarded as approximately pure oblate and prolate configurations.
Similar underestimation of the mixing has also been reported in the shell-model calculation~\cite{Frycz2024}. This observation motivates a different strategy. Rather than predicting the mixing amplitudes from the effective interaction, we use the $0_{\rm ob}^+$, $2_{\rm ob}^+$, $0_{\rm pr}^+$, and $2_{\rm pr}^+$ states as basis states and constrain their mixing amplitudes directly from the available experimental observables, as discussed in the next section.

The charge radius provides another important constraint.
The calculated charge radii of both the $0^+_{\rm ob}$ and $0^+_{\rm pr}$ states are larger than the observed charge radius of the ground state.
Therefore, the experimental radius cannot be reproduced by only mixing these two configurations.
Similar overestimations of charge radii have been reported systematically in HFB calculations for $sd$-shell nuclei~\cite{Delaroche2010} using the Gogny interaction. 
This systematic trend suggests a deficiency in the density-dependent term of the Gogny interaction, which is introduced to reproduce the saturation density of nuclei.
In light nuclei such as $^{28}$Si, the central density often exceeds the saturation density~\cite{Foris1987}, resulting in relatively smaller radii compared with the trend of heavier mass nuclei.
To account for this, we introduce a scaling factor $c_3$ for the density-dependent term, as explained above, and slightly reduce its strength from the original D1S parameter set to obtain the smaller charge radii.
The allowed range of $c_3$ will be constrained simultaneously with the mixing amplitudes using the experimental charge radius.

\section{Constraint on the mixing amplitudes}\label{sec:constraint}
In this section, we determine the allowed values of the oblate–prolate mixing amplitudes and the scaling factor $c_3$.
As explained in the previous section, the mixing amplitudes are treated as phenomenological parameters, while $c_3$ is varied to account for the charge radius.

For each fixed value of $c_3$, the oblate- and prolate-deformed basis states, $0^+_{\rm ob}$, $2^+_{\rm ob}$, $0^+_{\rm pr}$, and $2^+_{\rm pr}$, are generated by the AMD+GCM calculation, and the mixed states $0^+_1$, $0^+_3$, and $2^+_1$ are constructed by varying the two mixing angles $\theta_0$ and $\theta_2$.
\begin{align}
 &\ket{0_1^+} = \cos\theta_0 \ket{0_\ob^+} + \sin\theta_0\ket{0_\pr^+}, \label{eq:mix01}\\
 &\ket{0_3^+} = -\sin\theta_0 \ket{0_\ob^+} + \cos\theta_0\ket{0_\pr^+},\label{eq:mix03}\\
 &\ket{2_1^+} = \cos\theta_2 \ket{2_\ob^+} + \sin\theta_2\ket{2_\pr^+}. \label{eq:mix2}
\end{align}
The relative phases of the basis states are chosen such that the E2 matrix elements,
$M(E2;2^+_{\rm ob}\to 0^+_{\rm ob})$ and
$M(E2;2^+_{\rm pr}\to 0^+_{\rm pr})$, are both positive.
The allowed region in the $(\cos^2\theta_0,\cos^2\theta_2)$ plane is then obtained by requiring simultaneous reproduction of the charge radius, the quadrupole moment of the $2^+_1$ state, and the in-band and inter-band $B(E2)$ values.
Repeating this analysis for different values of $c_3$ gives the allowed region in the full $(\cos^2\theta_0,\cos^2\theta_2,c_3)$ parameter space.

\begin{figure}[tbp]
    \centering
    \includegraphics[width=\linewidth]{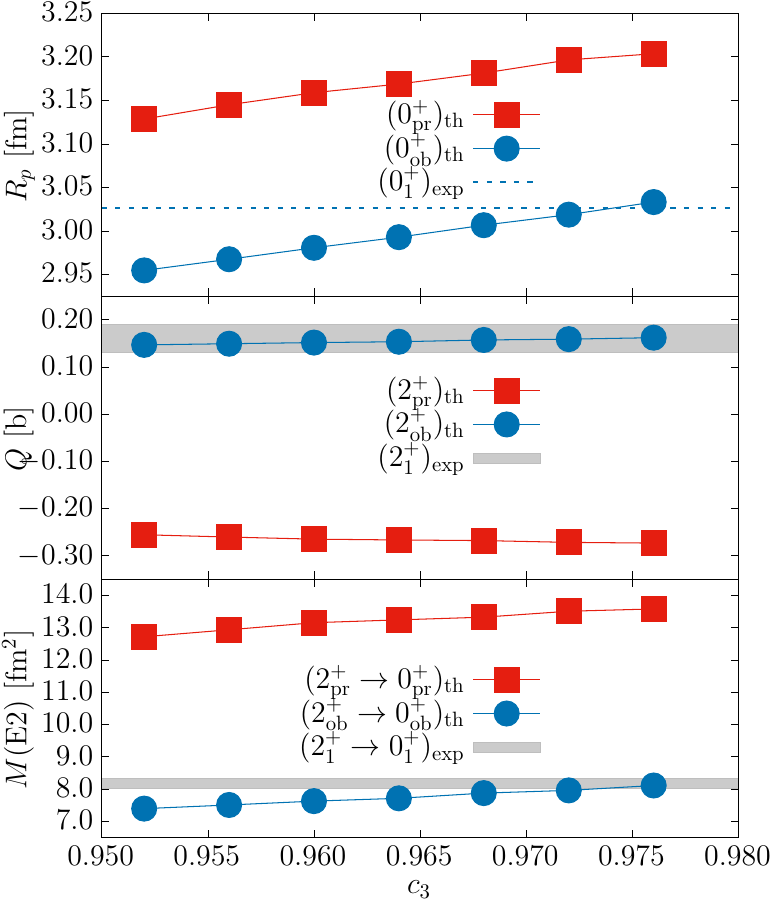}
    \caption{
    Proton radius (upper), Q-moments (middle), and E2 matrix elements (lower) as functions of $\coeffdep$. 
    In all panels, circles and squares represent the oblate and prolate configurations, respectively.
    For the theoretical Q-moment for the prolate configuration, the sign is inverted. 
    The dotted line shows the experimental proton radius.
    In the middle and lower panels, the shaded area shows the experimental values with the uncertainties.
    }
    \label{fig:e0e2_mat_ele}
\end{figure}

Figure~\ref{fig:e0e2_mat_ele} shows the $c_3$ dependence of the quantities used for the constraints, which  are the proton radii of the $0^+$ basis states, the quadrupole moments of the
$2^+$ basis states, and the E2 transition matrix elements within the oblate and prolate basis states.
We first consider the constraint from the charge radius.
As shown in the upper panel of Fig.~\ref{fig:e0e2_mat_ele}, the proton radius of the prolate configuration is larger than that of the oblate configuration for all values of $c_3$.
For a fixed value of $c_3$, the proton radius of the mixed ground state is given by
\begin{align}
 R_p^2({0}_1^+)  =
 R_p^2(0_{\rm ob}^+)\cos^2\theta_0  +  R_p^2(0_{\rm pr}^+)\sin^2\theta_0 .
 \label{eq:radius_constraint}
\end{align}
Thus, the experimental radius can be reproduced only when it lies between the radii of the oblate and prolate basis states. This condition already gives a rough constraint on $c_3$. 
For $c_3\gtrsim 0.975$, both the oblate and prolate radii are larger than the experimental value, and the observed radius cannot be reproduced.
A lower bound of $c_3$ is obtained by requiring that the ground state is oblate dominant. 
For $c_3 \lesssim 0.95$, the experimental radius cannot be reproduced with the oblate-dominant condition,
$\cos^2\theta_0 \geq 1/2$. Thus, $c_3$ is roughly constrained within the range of $0.95 \lesssim c_3 \lesssim 0.975$. 

\begin{figure}[tbp]
    \centering
    \includegraphics[width=\linewidth]{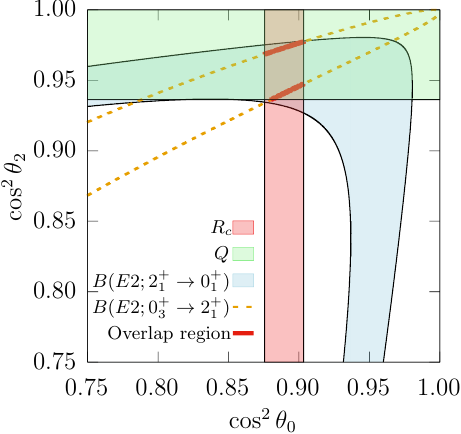}
    \caption{
    Constraints in the $(\cos^2\theta_0, \cos^2\theta_2)$ parameter space within experimental uncertainties for $\coeffdep = 0.968$. 
    The light red, green, blue, and yellow bands show the constraints from the proton radius $R_p$, the Q-moment $Q(2_1^+)$, $B(E2; 2_1^+ \rightarrow 0_1^+)$, and $B(E2; 0_3^+ \rightarrow 2_1^+)$, respectively. 
    The dark red region indicates the overlap of these constraints.
    }
    \label{fig:t3_968}
\end{figure}

For each fixed value of $c_3$, Eq.~(\ref{eq:radius_constraint}) gives the allowed range of $\cos^2\theta_0$.
As an example, the red band in Fig.~\ref{fig:t3_968} shows this radius constraint on the $(\cos^2\theta_0,\cos^2\theta_2)$ plane for $c_3=0.968$. Note that the position of this band changes with $c_3$, because the radii of the oblate and prolate basis states depend on $c_3$.

We next consider the constraint from the quadrupole moment of the $2^+_1$ state shown in the middle panel of Fig.~\ref{fig:e0e2_mat_ele}. The oblate and prolate basis states have different signs of the quadrupole moments, reflecting their opposite intrinsic deformations.
Similarly to the proton radius, the quadrupole moment of the mixed $2^+_1$ state is written as
\begin{align}
 Q(2_1^+)  =  Q(2_{\rm ob}^+)\cos^2\theta_2  +  Q(2_{\rm pr}^+)\sin^2\theta_2 .
 \label{eq:q_constraint}
\end{align}
Thus, the quadrupole moment gives a constraint on the mixing angle $\theta_2$.
Since the experimental value is positive and consistent with the oblate configuration, an oblate-dominant $2^+_1$ state is favored. The green band in Fig.~\ref{fig:t3_968} shows this constraint for the $c_3=0.968$ case.

We then consider the constraint from the E2 transition strength
$B(E2;2^+_1\to 0^+_1)$.
The lower panel of Fig.~\ref{fig:e0e2_mat_ele} shows the E2 transition
matrix elements within the oblate and prolate basis states.
Neglecting the small E2 matrix elements between the oblate and prolate
basis states, the E2 matrix element is written as
\begin{align}
 M(E2;2_1^+\to 0_1^+)
 =
 M(E2;2_{\rm ob}^+\to0_{\rm ob}^+)\cos\theta_0\cos\theta_2 \nonumber \\
 + M(E2;2_{\rm pr}^+\to0_{\rm pr}^+)\sin\theta_0\sin\theta_2 .
 \label{eq:be2_inband_constraint}
\end{align}
In principle, the experimental $B(E2)$ value determines only the magnitude of the left-hand side of this equation. However, because of the oblate dominance in the $0^+_1$ and $2^+_1$ states, the first term on the right-hand side should be much larger than the second term. Thus, we take $M(E2;2^+_1\to0^+_1)$ to be positive and obtain the blue band in Fig.~\ref{fig:t3_968} as the constraint on the
$(\cos^2\theta_0,\cos^2\theta_2)$ plane.

We finally consider the constraint from the inter-band transition strength $B(E2;0^+_3\to 2^+_1)$.
Using the same approximation as above, the corresponding E2 matrix element is written as
\begin{align}
 M(E2;0_3^+\to 2_1^+)
 =
 M(E2;0_{\rm pr}^+\to2_{\rm pr}^+)\cos\theta_0\sin\theta_2 \nonumber \\
 - M(E2;0_{\rm ob}^+\to2_{\rm ob}^+)\sin\theta_0\cos\theta_2,
 \label{eq:be2_interband_constraint}
\end{align}
which shows that the inter-band transition arises from the interference between the two configurations. In this case, the first and second terms on the right-hand side can have the same order of magnitude, and hence we take both signs of $M(E2;0^+_3\to2^+_1)$. The resulting constraints are shown by the yellow lines in Fig.~\ref{fig:t3_968}.

\begin{figure}[tbp]
    \centering
   \includegraphics[width=\linewidth]{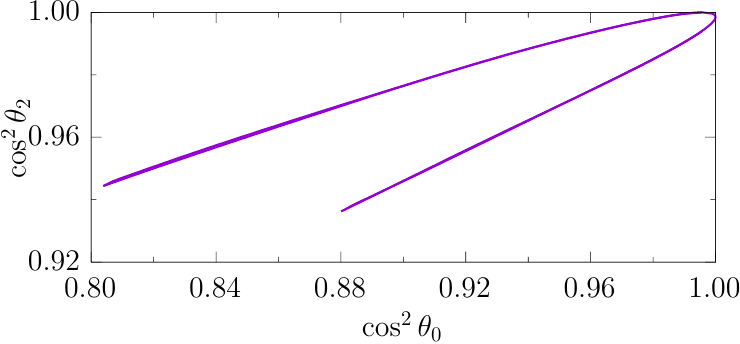}
    \caption{Overlap regions of the constraints in the $(\cos^2\theta_0, \cos^2\theta_2)$ parameter space over the full allowed range of $\coeffdep$.}
    \label{fig:mixing_area_all}
\end{figure}

By taking the overlap of the four constraints discussed above, we obtain the constraint for
the $c_3=0.968$ case, as shown by the dark-red region in Fig.~\ref{fig:t3_968}.
Repeating the same analysis for different values of $c_3$, we find a nonempty overlap in the range
$c_3=0.96$--$0.98$. Figure~\ref{fig:mixing_area_all} shows the projection of the resulting allowed region onto the $(\cos^2\theta_0,\cos^2\theta_2)$ plane.

The combined constraints give $\cos^2\theta_0 \gtrsim 0.8$, which corresponds to a prolate component of the ground state of less than about 20\%. 
For the $2^+_1$ state, a prolate component is less than 7\%.

Finally, we estimate the E0 transition strength between the two mixed
$0^+$ states as a consequence of the allowed mixing amplitudes.
Again, assuming the E0 matrix element between the oblate and prolate basis
states is negligible, the inter-band E0 matrix element is given as
\begin{equation}
  M(\mathrm{E0}; 0_3^+ \rightarrow 0_1^+) 
  = Z \left[ R_p^2\left(0_\pr^+\right) - R_p^2(0_\ob^+) \right]
  \cos\theta_0\sin\theta_0,
  \label{e0inter}
\end{equation}
where $Z=14$ is the proton number. Note that $R_p^2\left(0_\pr^+\right) - R_p^2(0_\ob^+)$
is almost independent of the value of $c_3$, as seen in Fig.~\ref{fig:e0e2_mat_ele}.
Therefore, the E0 strength is sensitive to the amount of mixing in the $0^+$ state.
\begin{figure}[tbp]
    \centering
   \includegraphics[width=\linewidth]{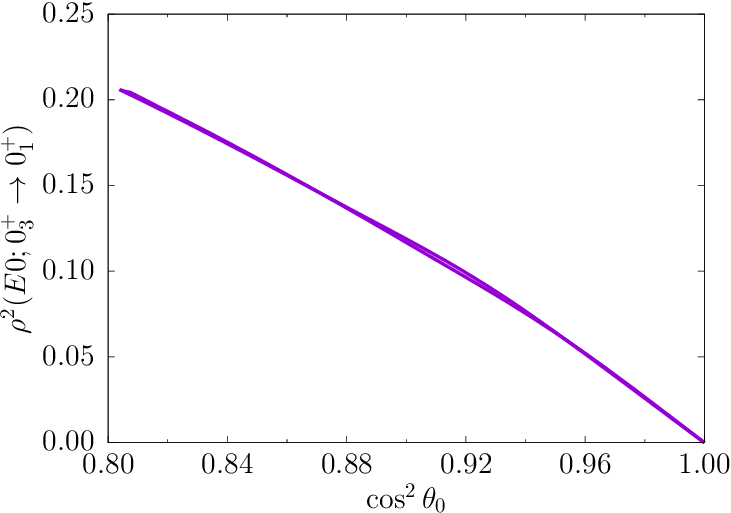}
    \caption{Predicted dimensionless E0 transition strength of the $0_3^+ \to 0_1^+$ transition. 
    }
    \label{fig:be0}
\end{figure}
Figure~\ref{fig:be0} shows the estimated value of 
$\rho^2(\mathrm{E0};0_3^+\to0_1^+)$ obtained from the allowed parameter region.
The estimated strength covers a broad range, approximately
$\rho^2(\mathrm{E0};0_3^+\to0_1^+) \lesssim 0.20$, 
and hence the present constraints do not provide a sharp prediction for the E0 transition strength.
Therefore, an experimental E0 strength would provide a useful test of the simple two-state mixing picture discussed here.

\section{Conclusions}\label{sec:conclusion}
We have investigated oblate-prolate shape mixing in the low-lying states
of $\nucl{28}{Si}$ by constructing mixed states from the oblate- and
prolate-dominant AMD+GCM wave functions.
The mixing amplitudes were introduced as phenomenological parameters and constrained
using the available experimental observables: the charge radius, quadrupole moment, and in-band and inter-band E2 transition strengths.
The strength of the density-dependent term in the Gogny D1S interaction
was also varied to account for the charge radius.

The combined constraints give $\cos^2\theta_0 \gtrsim 0.8$, corresponding
to a prolate component of less than about 20\% in the ground state.
For the $2_1^+$ state, the allowed prolate component is even smaller.
We also estimated the E0 transition strength between the two mixed
$0^+$ states from the allowed parameter region.
The present constraints do not determine the E0 strength sharply, but they
provide an upper limit of about
$\rho^2(\mathrm{E0};0_3^+\to0_1^+) \lesssim 0.20$.

\section*{Acknowledgment}
This work was supported by JSPS KAKENHI Grant Nos.~22K03610, 23K22485, and 26K00703; by JSPS Bilateral Program Number JPJSBP120247715 and the Department of Science and Technology (DST), Government of India, No.~DST/INT/JSPS/P-393/2024(G); and by the Multidisciplinary Cooperative Research Program in CCS, University of Tsukuba (MCRP).
Numerical calculations were performed on computing resources provided through the MCRP and at the Yukawa Institute Computer Facility.

\bibliography{28Si_Shape_mixing}

\end{document}